\documentclass[12pt,preprint]{aastex62}

\usepackage{amsmath,amsthm}
\received{\today}
\revised{\today}
\accepted{\today}
\submitjournal{ApJ}

\begin{document}

\title{Measurement of the extragalactic background light spectral energy distribution with VERITAS}

\correspondingauthor{Elisa Pueschel}
\email{elisa.pueschel@desy.de}
\author{A.~U.~Abeysekara}\affiliation{Department of Physics and Astronomy, University of Utah, Salt Lake City, UT 84112, USA}
\author{A.~Archer}\affiliation{Department of Physics and Astronomy, DePauw University, Greencastle, IN 46135-0037, USA}
\author{W.~Benbow}\affiliation{Center for Astrophysics $|$ Harvard \& Smithsonian, Cambridge, MA 02138, USA}
\author{R.~Bird}\affiliation{Department of Physics and Astronomy, University of California, Los Angeles, CA 90095, USA}
\author{A.~Brill}\affiliation{Physics Department, Columbia University, New York, NY 10027, USA}
\author{R.~Brose}\affiliation{Institute of Physics and Astronomy, University of Potsdam, 14476 Potsdam-Golm, Germany and DESY, Platanenallee 6, 15738 Zeuthen, Germany}
\author{M.~Buchovecky}\affiliation{Department of Physics and Astronomy, University of California, Los Angeles, CA 90095, USA}
\author{D.~Calderon-Madera}\affiliation{Department of Physics, California State University - East Bay, Hayward, CA 94542, USA}
\author{J.~L.~Christiansen}\affiliation{Physics Department, California Polytechnic State University, San Luis Obispo, CA 94307, USA}
\author{W.~Cui}\affiliation{Department of Physics and Astronomy, Purdue University, West Lafayette, IN 47907, USA and Department of Physics and Center for Astrophysics, Tsinghua University, Beijing 100084, China.}
\author{M.~K.~Daniel}\affiliation{Center for Astrophysics $|$ Harvard \& Smithsonian, Cambridge, MA 02138, USA}
\author{A.~Falcone}\affiliation{Department of Astronomy and Astrophysics, 525 Davey Lab, Pennsylvania State University, University Park, PA 16802, USA}
\author{Q.~Feng}\affiliation{Physics Department, Columbia University, New York, NY 10027, USA}
\author{M.~Fernandez-Alonso}\affiliation{Instituto de Astronom'a y F'sica del Espacio (IAFE, CONICET-UBA), CC 67 - Suc. 28, (C1428ZAA) Ciudad Aut—noma de Buenos Aires, Argentina}
\author{J.~P.~Finley}\affiliation{Department of Physics and Astronomy, Purdue University, West Lafayette, IN 47907, USA}
\author{L.~Fortson}\affiliation{School of Physics and Astronomy, University of Minnesota, Minneapolis, MN 55455, USA}
\author{A.~Furniss}\affiliation{Department of Physics, California State University - East Bay, Hayward, CA 94542, USA}
\author{A.~Gent}\affiliation{School of Physics and Center for Relativistic Astrophysics, Georgia Institute of Technology, 837 State Street NW, Atlanta, GA 30332-0430}
\author{C.~Giuri}\affiliation{DESY, Platanenallee 6, 15738 Zeuthen, Germany}
\author{O.~Gueta}\affiliation{DESY, Platanenallee 6, 15738 Zeuthen, Germany}
\author{D.~Hanna}\affiliation{Physics Department, McGill University, Montreal, QC H3A 2T8, Canada}
\author{T.~Hassan}\affiliation{DESY, Platanenallee 6, 15738 Zeuthen, Germany}
\author{O.~Hervet}\affiliation{Santa Cruz Institute for Particle Physics and Department of Physics, University of California, Santa Cruz, CA 95064, USA}
\author{J.~Holder}\affiliation{Department of Physics and Astronomy and the Bartol Research Institute, University of Delaware, Newark, DE 19716, USA}
\author{G.~Hughes}\affiliation{Center for Astrophysics $|$ Harvard \& Smithsonian, Cambridge, MA 02138, USA}
\author{T.~B.~Humensky}\affiliation{Physics Department, Columbia University, New York, NY 10027, USA}
\author{C.~A.~Johnson}\affiliation{Santa Cruz Institute for Particle Physics and Department of Physics, University of California, Santa Cruz, CA 95064, USA, now at NextEra Analytics, Inc., St. Paul, MN 55107, USA}
\author{P.~Kaaret}\affiliation{Department of Physics and Astronomy, University of Iowa, Van Allen Hall, Iowa City, IA 52242, USA}
\author{M.~Kertzman}\affiliation{Department of Physics and Astronomy, DePauw University, Greencastle, IN 46135-0037, USA}
\author{D.~Kieda}\affiliation{Department of Physics and Astronomy, University of Utah, Salt Lake City, UT 84112, USA}
\author{M.~Krause}\affiliation{DESY, Platanenallee 6, 15738 Zeuthen, Germany}
\author{F.~Krennrich}\affiliation{Department of Physics and Astronomy, Iowa State University, Ames, IA 50011, USA}
\author{S.~Kumar}\affiliation{Physics Department, McGill University, Montreal, QC H3A 2T8, Canada}
\author{M.~J.~Lang}\affiliation{School of Physics, National University of Ireland Galway, University Road, Galway, Ireland}
\author{G.~Maier}\affiliation{DESY, Platanenallee 6, 15738 Zeuthen, Germany}
\author{P.~Moriarty}\affiliation{School of Physics, National University of Ireland Galway, University Road, Galway, Ireland}
\author{R.~Mukherjee}\affiliation{Department of Physics and Astronomy, Barnard College, Columbia University, NY 10027, USA}
\author{M.~Nievas-Rosillo}\affiliation{DESY, Platanenallee 6, 15738 Zeuthen, Germany}
\author{S.~O'Brien}\affiliation{Physics Department, McGill University, Montreal, QC H3A 2T8, Canada}
\author{R.~A.~Ong}\affiliation{Department of Physics and Astronomy, University of California, Los Angeles, CA 90095, USA}
\author{K.~Pfrang}\affiliation{DESY, Platanenallee 6, 15738 Zeuthen, Germany}
\author{M.~Pohl}\affiliation{Institute of Physics and Astronomy, University of Potsdam, 14476 Potsdam-Golm, Germany and DESY, Platanenallee 6, 15738 Zeuthen, Germany}
\author{R.~R.~Prado}\affiliation{DESY, Platanenallee 6, 15738 Zeuthen, Germany}
\author{E.~Pueschel}\affiliation{DESY, Platanenallee 6, 15738 Zeuthen, Germany}
\author{J.~Quinn}\affiliation{School of Physics, University College Dublin, Belfield, Dublin 4, Ireland}
\author{K.~Ragan}\affiliation{Physics Department, McGill University, Montreal, QC H3A 2T8, Canada}
\author{P.~T.~Reynolds}\affiliation{Department of Physical Sciences, Cork Institute of Technology, Bishopstown, Cork, Ireland}
\author{D.~Ribeiro}\affiliation{Physics Department, Columbia University, New York, NY 10027, USA}
\author{G.~T.~Richards}\affiliation{Department of Physics and Astronomy and the Bartol Research Institute, University of Delaware, Newark, DE 19716, USA}
\author{E.~Roache}\affiliation{Center for Astrophysics $|$ Harvard \& Smithsonian, Cambridge, MA 02138, USA}
\author{A.~C.~Rovero}\affiliation{Instituto de Astronom'a y F'sica del Espacio (IAFE, CONICET-UBA), CC 67 - Suc. 28, (C1428ZAA) Ciudad Aut—noma de Buenos Aires, Argentina}
\author{I.~Sadeh}\affiliation{DESY, Platanenallee 6, 15738 Zeuthen, Germany}
\author{M.~Santander}\affiliation{Department of Physics and Astronomy, University of Alabama, Tuscaloosa, AL 35487, USA}
\author{G.~H.~Sembroski}\affiliation{Department of Physics and Astronomy, Purdue University, West Lafayette, IN 47907, USA}
\author{K.~Shahinyan}\affiliation{School of Physics and Astronomy, University of Minnesota, Minneapolis, MN 55455, USA}
\author{I.~Sushch}\affiliation{Institute of Physics and Astronomy, University of Potsdam, 14476 Potsdam-Golm, Germany}
\author{T.~Svraka}\affiliation{Department of Physics, California State University - East Bay, Hayward, CA 94542, USA}
\author{A.~Weinstein}\affiliation{Department of Physics and Astronomy, Iowa State University, Ames, IA 50011, USA}
\author{R.~M.~Wells}\affiliation{Department of Physics and Astronomy, Iowa State University, Ames, IA 50011, USA}
\author{P.~Wilcox}\affiliation{Department of Physics and Astronomy, University of Iowa, Van Allen Hall, Iowa City, IA 52242, USA}
\author{A.~Wilhelm}\affiliation{Institute of Physics and Astronomy, University of Potsdam, 14476 Potsdam-Golm, Germany and DESY, Platanenallee 6, 15738 Zeuthen, Germany}
\author{D.~A.~Williams}\affiliation{Santa Cruz Institute for Particle Physics and Department of Physics, University of California, Santa Cruz, CA 95064, USA}
\author{T.~J~Williamson}\affiliation{Department of Physics and Astronomy and the Bartol Research Institute, University of Delaware, Newark, DE 19716, USA}
\author{B.~Zitzer}\affiliation{Physics Department, McGill University, Montreal, QC H3A 2T8, Canada}

%\author{Amy Hendrickson}
%\altaffiliation{Creator of AASTeX v6.2}
%\affiliation{TeXnology Inc.}
%\collaboration{(LaTeX collaboration)}

%% Note that the \and command from previous versions of AASTeX is now
%% depreciated in this version as it is no longer necessary. AASTeX 
%% automatically takes care of all commas and "and"s between authors names.

%% AASTeX 6.2 has the new \collaboration and \nocollaboration commands to
%% provide the collaboration status of a group of authors. These commands 
%% can be used either before or after the list of corresponding authors. The
%% argument for \collaboration is the collaboration identifier. Authors are
%% encouraged to surround collaboration identifiers with ()s. The 
%% \nocollaboration command takes no argument and exists to indicate that
%% the nearby authors are not part of surrounding collaborations.

%% Mark off the abstract in the ``abstract'' environment. 
\begin{abstract}

The extragalactic background light (EBL), a diffuse photon field in the optical and infrared range, is a record of radiative processes over the Universe's history. Spectral measurements of blazars at very high energies ($>$100 GeV) enable the reconstruction of the spectral energy distribution (SED) of the EBL, as the blazar spectra are modified by redshift- and energy-dependent interactions of the gamma-ray photons with the EBL. The spectra of 14 VERITAS-detected blazars are included in a new measurement of the EBL SED that is independent of EBL SED models. The resulting SED covers an EBL wavelength range of 0.56--56 $\mu$m, and is in good agreement with lower limits obtained by assuming that the EBL is entirely due to radiation from cataloged galaxies. 

\end{abstract}

%% Keywords should appear after the \end{abstract} command. 
%% See the online documentation for the full list of available subject
%% keywords and the rules for their use.

%% From the front matter, we move on to the body of the paper.
%% Sections are demarcated by \section and \subsection, respectively.
%% Observe the use of the LaTeX \label
%% command after the \subsection to give a symbolic KEY to the
%% subsection for cross-referencing in a \ref command.
%% You can use LaTeX's \ref and \label commands to keep track of
%% cross-references to sections, equations, tables, and figures.
%% That way, if you change the order of any elements, LaTeX will
%% automatically renumber them.
%%
%% We recommend that authors also use the natbib \citep
%% and \citet commands to identify citations.  The citations are
%% tied to the reference list via symbolic KEYs. The KEY corresponds
%% to the KEY in the \bibitem in the reference list below. 

\section{Introduction} \label{sec:intro}
The Universe is permeated by diffuse photon fields, tracking light emission from the epoch of recombination to the present day. At ultraviolet (UV), optical and infrared wavelengths, the extragalactic background light (EBL) is primarily due to stellar radiation in the UV/optical band, and re-radiation of starlight into the infrared band after absorption by matter (primarily dust). Secondary contributions to the UV/optical intensity come from energy release associated with accretion of matter in active galactic nuclei. Further, more exotic contributions have been hypothesized, including dark matter decay in the early universe and radiation from primordial stars~\citep{Dwek2005}. A number of comprehensive reviews discuss the EBL~\citep{Hauser2001, Krennrich2013}; only the key points are discussed here.

The EBL intensity as a function of wavelength and its evolution with redshift have been predicted by a number of authors~\citep{Franceschini2017, Franceschini2018, Gilmore2012,Finke2010, Dominguez2011}. While a detailed description of these models is beyond the scope of this paper, these approaches require information about stellar and galactic formation and evolution, and models of the matter (dust) distribution in galaxies and of radiative transfer. Comparing the measured spectral energy distribution (SED) of the EBL against theoretical predictions thus tests the current understanding of the density of and radiation by visible matter since the epoch of reionization.

While direct measurements of the EBL SED have been attempted, most recently by \cite{CIBER} and \cite{NewHorizons}, correct subtraction of contaminating foregrounds is challenging, and the measurements tend to be limited by systematic uncertainties. Robust lower bounds on the EBL SED, on the other hand, have been generated by summing the light expected based on resolved galaxies. Deep surveys in recent years have improved the precision of these measurements~\citep{Madau2000, Fazio2004, Keenan2010, Driver2016}. It is worth noting that by design, most EBL models are in good agreement with the limits from galaxy counts measurements. However, this approach is by construction insensitive to contributions from unresolved sources and diffuse emission. 

Spectral measurements of distant gamma-ray emitters provide an alternative method to measure the EBL SED, with an independent set of uncertainties and sensitivity to diffuse contributions. The interaction of very-high-energy (VHE; $>$100~GeV) gamma rays with EBL photons has been discussed since the early days of VHE gamma-ray astronomy~\citep{Nikishov1962, Gould1967}. At center-of-mass energies above the energy threshold for electron-positron pair creation, the VHE gamma-ray flux will be attenuated by interaction with the EBL. The probability of an interaction increases with VHE gamma-ray energy and with the distance the gamma-ray traverses. Consequently, EBL attenuation results in differences between the EBL-absorbed and intrinsic spectra of VHE gamma-ray emitters. For a given energy and source distance, these differences are expected to evolve in the same way in blazar spectra, independent of the source. 
  
High-frequency-peaked blazars (HBLs) are canonically used in measurements of EBL-induced spectral modification, owing to their abundance ($\sim$50 HBLs were detected in the VHE range at the time of writing\footnote{http://tevcat.uchicago.edu}), their detection at VHE to large redshifts (for example, $z$=0.604 for PKS 1424+240), and their emission of gamma rays up to high energies (detected to tens of TeV for nearby sources). Intermediate-frequency-peaked blazars (IBLs), though less abundant in the VHE catalog, can also be used in EBL measurements. As discussed in Section~\ref{sec:Analysis}, it is necessary to make some assumptions about the intrinsic source spectra to extract the EBL imprint from the observed spectra. However, biases resulting from such choices are mitigated by making conservative assumptions and utilizing observations of a large sample of HBLs and IBLs located at a range of distances.

\section{VERITAS} \label{sec:VERITAS}
VERITAS utilizes the imaging atmospheric Cherenkov technique, measuring Cherenkov emission induced by particle showers in Earth's atmosphere. The imaging of such showers allows the reconstruction of the energy and arrival direction of their gamma-ray progenitors. VERITAS consists of an array of four telescopes~\citep{VERITASinstrument} located at the Fred Lawrence Whipple Observatory in southern Arizona, USA (+31$^{\circ}$ 40$\arcmin$30.21$\arcsec$, --110$^{\circ}$ 57$\arcmin$ 7.77$\arcsec$, 1268 m above sea level). Each telescope includes a 12 meter reflector of Davies-Cotton design~\citep{Davies1957} and a photomultiplier tube camera with 499 pixels. The instrument has been upgraded several times: in 2009, one of the telescopes was relocated, improving sensitivity with a more symmetric array, and in 2012, the cameras and trigger system were upgraded, improving sensitivity at the lowest energies~\citep{Kieda2013}. The instrument is sensitive to gamma rays with energies from $\sim$85 GeV to greater than 30 TeV, has a field of view of 3.5$^{\circ}$, an energy resolution of 15--25\%, and an angular resolution (given as the 68\% containment radius) of $<$0.1$^{\circ}$ at 1 TeV~\citep{VERITASspecs}. 

The data used in this measurement were collected with the camera center offset by 0.5$^{\circ}$ from the source position (\textit{wobble} pointing mode). This observation mode facilitates background estimation, as source and background regions can be defined within the same field of view~\citep{Fomin}. The data included in this analysis span from the beginning of VERITAS operations in September 2007 to June 2016. Data were collected under good weather conditions, and include observations taken under both dark sky conditions and, for a fraction of the data ($\sim$20\%), under moonlight~\citep{moonpaper}. The increased energy threshold of moonlight observations was considered acceptable, as the spectral behavior at the highest energies is of interest for EBL measurements. The data were processed with one of the standard VERITAS calibration and event reconstruction pipelines~\citep{GernotICRC2017, BDTpaper}, and the results were cross-checked with an independent chain~\citep{VEGAS, VERITASanalysisgen1, VERITASanalysisgen2}. A gamma/hadron selection corresponding to an energy threshold (defined by the average energy bias falling below 10\%) of $\sim$170~GeV (205~GeV for a few cases) was used to preserve sensitivity to low-energy gamma rays. 

\section{Source analysis}\label{sec:SourceAnalysis}
The blazars considered in this analysis are listed in Table~\ref{sourcelist}. The source sample comprises thirteen HBLs and one intermediate-frequency peaked blazar (IBL), 3C 66A. The source sample was selected with the goal of smoothly covering a broad redshift range, while focusing on objects with high detection significance to minimize statistical uncertainties. As discussed in Section~\ref{sec:Analysis}, the photon spectrum is the primary input to the EBL analysis, and the uncertainties on the spectral points determine the uncertainty of the EBL measurement. 

\begin{table}
\centerline{
\begin{tabular}{cccccc}
Target & Redshift & Exposure [min] & $\sigma_{\textrm{detect}}$ & Flux variability & Spectral variability \\
\hline
\hline
1ES 2344+514  & 0.044 & 4514 & 31.0 & yes & no \\
1ES 1959+650 & 0.048 & 3324 & 102.5 & yes & yes \\
RGB J0710+591 & 0.125 & 7926 & 11.5 & yes & no \\ 
H 1426+428 & 0.129 & 5477 & 13.3 & yes & no \\
1ES 1215+303 & 0.13 & 10071 & 33.0  & yes & no \\
1ES 0229+200 & 0.14 & 8392 & 12.3 & yes & no \\
1ES 1218+304 & 0.182 & 9524 & 63.0 & yes & no \\
1ES 1011+496 & 0.212  & 2160 & 43.5 & yes & no \\
MS 1221.8+2452 & 0.218 & 152 & 22.0 & yes & no \\
1ES 0414+009 & 0.287 & 6457 & 9.3 & no & no \\
1ES 0502+675 & 0.341 & 1970 & 13.9 & yes & no \\
3C 66A & 0.34--0.41 & 5926 & 26.4  & yes & yes \\
PG 1553+113 & 0.43--0.58 & 7774 & 71.8 & yes & no \\
PKS 1424+240 & 0.604  & 10697 & 28.3 & yes & no \\
 \hline
\end{tabular}
}
\caption{\small{Targets analyzed, ordered by redshift. The VERITAS observing time (before deadtime correction) is given in the third column, the detection significance in the fourth column, and whether or not the source shows flux and spectral variability in the fifth and sixth columns, respectively.}}
\label{sourcelist}
\end{table}

A distinguishing characteristic of blazars is their flux variability, which in some cases has been shown to correlate with spectral variability~\citep{Albert2007, Fossati2008, Abramowski2010}. Averaging together hard and soft spectral shapes can distort the shape of the time-averaged spectrum, leading to the appearance of spectral hardening at high energy if a spectrally hard bright state and a spectrally soft dim state are averaged. To avoid this and the consequent risk of biases, the flux and spectral variability of the sources were studied. For sources that showed evidence for flux variability (as shown in Table~\ref{sourcelist}, all sources other than 1ES 0414+009), the datasets were divided by flux level and spectral fits were performed for each flux bin. The spectra for sources that showed possible spectral variability ($>$2$\sigma$ variation in a spectral parameter other than the flux normalization) are divided into high- and low-state spectra and treated independently in the EBL analysis. As shown in Table~\ref{sourcelist}, only 1ES 1959+650 and 3C 66A show evidence for spectral variability. For 1ES 0502+675 and MS 1221.8+2452, the sources experienced bright states outside of which the sources were not detected, for the former over a single observing period, for the latter, during a two night flare. For these two sources, spectra were constructed only from data collected during these bright periods. 

%A full listing of the differential flux points corresponding to the photon spectra is given in Appendix~\ref{appendixA}. 
Three spectral models are considered: a simple power law ($dN/dE = N_{0}(E/E_{0})^{-\Gamma}$), a power law with an exponential cut-off ($dN/dE = N_{0}(E/E_{0})^{-\Gamma}\exp(-E/E_{\textrm{c}})$), and a log parabola ($dN/dE = N_{0}(E/E_{0})^{-a-b\log(E/E_{0})}$). A binned $\chi^{2}$ fit is utilized. The majority of the spectra are well-described by a simple power law, in which case a more complex fit function is not considered. Table~\ref{powerlawobserved} summarizes the fit results for all observed spectra that are well described by a power law, while Table~\ref{eplobserved} and Table~\ref{logparabolaobserved} show the fit results for observed spectra best described by a power law with an exponential cut-off and a log parabola, respectively. For several cases, both a power law with exponential cutoff and a log parabola provided an acceptable fit, in which case the better fit is shown. 

\begin{table}
\centerline{
\begin{tabular}{ccccc}
Target   &  $E_{0}$ [TeV] & $N_{0}$ [cm$^{-2}$s$^{-1}$TeV$^{-1}$] & $\Gamma$ & $\chi^{2}/\textrm{ndof}$ \\
\hline
\hline
RGB J0710+591             & 0.7 & (1.37$\pm$0.13)$\times 10^{-12}$ & 2.85$\pm$0.12 & 3.45/6 \\
1ES 1215+303               & 0.7 & (1.02$\pm$0.08)$\times 10^{-12}$ & 3.67$\pm$0.09 & 8.16/6 \\
1ES 0229+200              & 0.7 & (8.24$\pm$0.65)$\times 10^{-13}$ & 3.01$\pm$0.10 & 10.93/6 \\
1ES 1218+304              & 0.7 & (2.70$\pm$0.09)$\times 10^{-12}$ & 3.40$\pm$0.04 & 8.33/5 \\
1ES 1011+496              & 0.7 & (4.23$\pm$0.26)$\times 10^{-12}$ & 3.68$\pm$0.07 & 8.92/4 \\
MS 1221.8+2452            & 0.7 & (6.71$\pm$0.95)$\times 10^{-12}$ & 3.31$\pm$0.12 & 4.18/4 \\
1ES 0414+009              & 0.3 & (5.65$\pm$0.72)$\times 10^{-12}$ & 4.08$\pm$0.16 & 10.87/5 \\
1ES 0502+675              & 0.7 & (2.41$\pm$0.29)$\times 10^{-12}$ & 3.80$\pm$0.25 & 0.99/3 \\         
 \hline
\end{tabular}
}
\caption{\small{Fit parameters for spectra that are well-described by a power law.}}
\label{powerlawobserved}
\end{table}

\begin{table}
\centerline{
\begin{tabular}{cccccc}
Target   &  $E_{0}$ [TeV] & $N_{0}$ [cm$^{-2}$s$^{-1}$TeV$^{-1}$] & $\Gamma$ & $E_{\textrm{c}}$ [TeV] & $\chi^{2}/\textrm{ndof}$ \\
\hline
\hline
1ES 1959+650 (high state) & 0.7 & (5.29$\pm$0.14)$\times 10^{-11}$ & 2.44$\pm$0.05 & 4.82$\pm$0.26 & 9.66/6 \\
 \hline
\end{tabular}
}
\caption{\small{Fit parameters for spectra that are well-described by a power law with an exponential cut-off.}}
\label{eplobserved}
\end{table}

\begin{table}
\centerline{
\begin{tabular}{cccccc}
Target   &  $E_{0}$ [TeV] & $N_{0}$ [cm$^{-2}$s$^{-1}$TeV$^{-1}$] & $a$ & $b$ & $\chi^{2}/\textrm{ndof}$ \\
\hline
\hline
1ES 2344+514               & 0.7 & (7.10$\pm$0.30)$\times 10^{-12}$ & 2.35$\pm$0.06 & 0.26$\pm$0.05 & 5.59/6 \\
1ES 1959+650 (low state)   & 0.7 & (9.74$\pm$0.74)$\times 10^{-12}$ & 2.63$\pm$0.16 & 0.30$\pm$0.06 & 3.30/5 \\
H 1426+428                 & 0.7 & (1.70$\pm$0.18)$\times 10^{-12}$ & 2.70$\pm$0.15 & 0.83$\pm$0.24 & 2.69/4 \\
3C 66A (high state)        & 0.7 & (1.37$\pm$0.63)$\times 10^{-12}$ & 6.41$\pm$0.93 & 1.38$\pm$0.44 & 1.12/2 \\
3C 66A (low state)         & 0.3 & (1.79$\pm$0.14)$\times 10^{-11}$ & 4.20$\pm$0.21 & 1.36$\pm$0.25 & 4.37/5 \\
PG 1553+113                & 0.3 & (3.03$\pm$0.10)$\times 10^{-13}$ & 4.78$\pm$0.11 & 1.35$\pm$0.18 & 6.81/4 \\
PKS 1424+240              & 0.1 & (6.82$\pm$2.29)$\times 10^{-10}$ & 3.83$\pm$0.90 & 0.82$\pm$0.51 & 2.19/4 \\
 \hline
\end{tabular}
}
\caption{\small{Fit parameters for spectra that are well-described by a log parabola.}}
\label{logparabolaobserved}
\end{table}

\section{EBL analysis}\label{sec:Analysis}
The approach taken here for extracting the EBL significance has been employed successfully in the literature~\citep{Mazin2007, Orr2011, Biteau2015}. Rather than evaluating the consistency of the observed blazar spectrum with a model (such as \citealt{Dominguez2011, Finke2010, Franceschini2008, Gilmore2012}), the gamma-ray extinction predicted by generic EBL shapes and its impact on the blazar spectra described in Section~\ref{sec:SourceAnalysis} are studied.

\begin{figure}
\centerline{\includegraphics[width=0.5\textwidth]{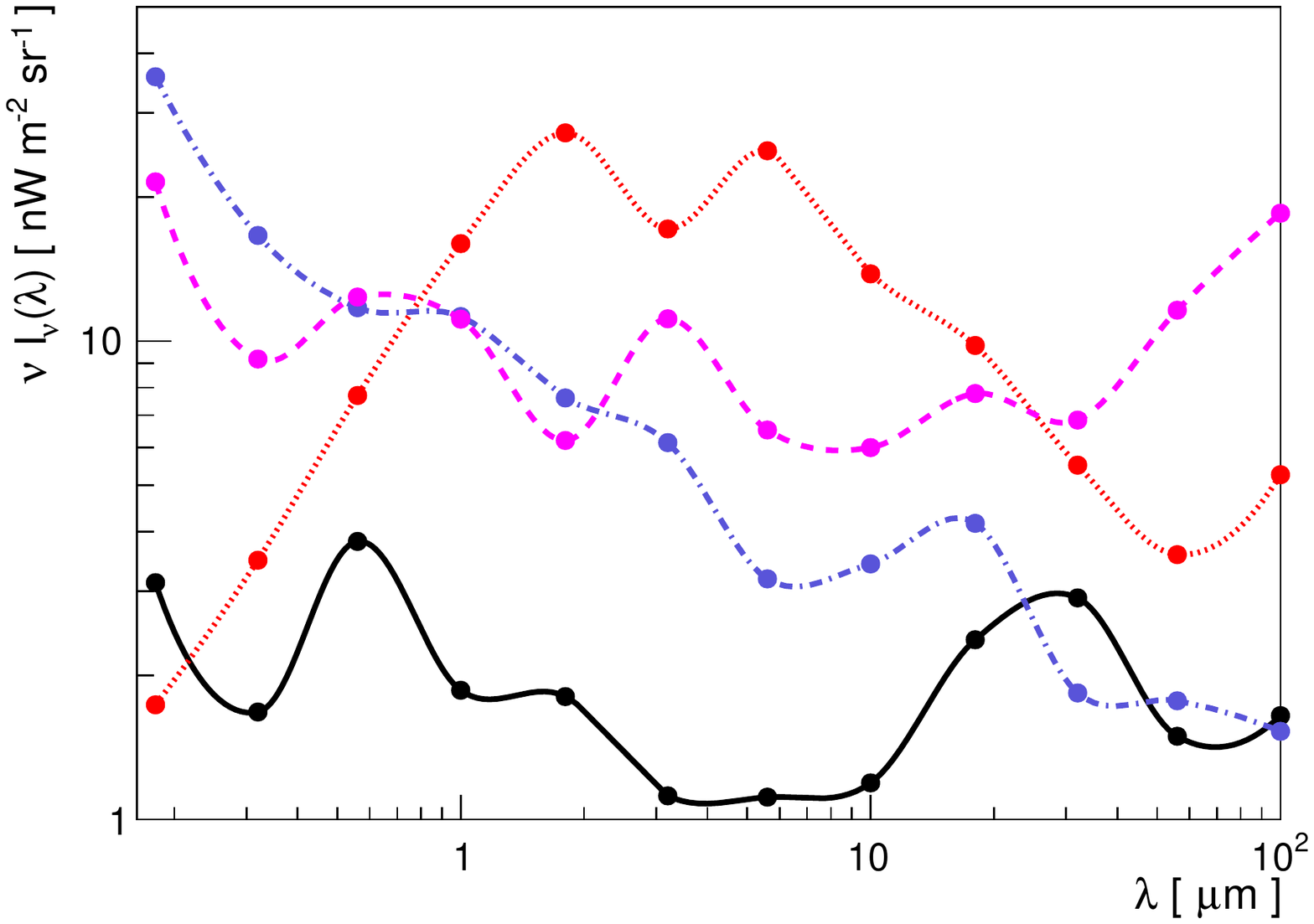}}
\caption{\small{Several examples of generic EBL shapes, defined as EBL intensity as a function of $\lambda_{\textrm{EBL}}$. The markers indicate grid points in $\lambda_{\textrm{EBL}}$ at which a random number is drawn to describe the EBL intensity.}}\label{fig:EBLshapes}
\end{figure}

An example of several EBL shapes is shown in Figure~\ref{fig:EBLshapes}. The shapes are generated by defining a set of 12 grid points in the wavelength of EBL photons, $\lambda_{\textrm{EBL}}$, within the range 0.1--100 $\mu$m. The distance between grid points is set to be of the order of the full width at half maximum of the EBL kernel (the pair-production cross section integrated along the line of sight). The EBL kernel width imposes a physical limit on the narrowest structure that can be probed. At each grid point, a value for the EBL intensity is set via random number generation between 1.0 and 50.0 nW~m$^{-2}$~sr$^{-1}$, producing a flat initial distribution in intensity. The intensity range was chosen to fully contain the upper and lower limits from complementary measurements. The 12 intensity values (one random value for each grid point in $\lambda_{\textrm{EBL}}$) define an EBL shape. The EBL intensity was not allowed to change by more than a factor of 2.5 between consecutive grid points, eliminating EBL shapes with unresolvably sharp structures. This restriction spoils the initially flat EBL intensity probability distribution. The flat starting distribution is regained by attaching a weight at each grid point for each EBL shape. This ensures that the results are not biased towards any particular EBL intensity due to a concentration of EBL shapes at that intensity. It should also be emphasized that external constraints from direct measurements or galaxy counts estimates are not considered in the generation of the EBL shapes. An ensemble of 480,000 EBL shapes is considered in the analysis described below, smoothly covering the intensity space from 1.0 to 50.0 nW~m$^{-2}$~sr$^{-1}$. 

The generated EBL shapes are taken to represent possible EBL SEDs at redshift $z$=0. They are represented as second order splines. As has been discussed in detail in the literature (e.g. \citealt{Krennrich2013, Biteau2015, HESS2017}), an EBL number density at $z$=0 can be translated into an optical depth $\tau(E, z)$, quantifying the attenuation of a gamma-ray flux of energy $E$ traveling a distance $z$. For each EBL shape $i$, $\tau_{i}(E, z)$ is numerically computed for a range of energies spanning 0.1--20.0 TeV and redshifts $z$=0.03--1.0. The energy and redshift points at which the calculations are made are spaced evenly in logarithmic space in energy, and semi-logarithmically in redshift. Flat $\Lambda$CDM cosmology is assumed in the calculation, with dark energy density $\Omega_{\Lambda}$ = 0.73, matter density $\Omega_{\textrm{M}}$ = 0.27, and Hubble constant $H_{0}$ = 70 km~s$^{-1}$~Mpc$^{-1}$. Fractional numerical integration errors are stored as $\sigma_{i}(\tau_{i}(E, z))$. Integration errors increase with energy and redshift, but on average are a few percent.

The evolution of the EBL with redshift must be considered; both the Universe's expansion and the evolution of radiation sources affect the intensity of the EBL. The evolution is handled empirically as proposed by \cite{Madau1996}: the EBL number density $\eta_{\textrm{EBL}}$ is scaled with redshift as $(1 + z)^{3 - f_{\textrm{evo}}}$ rather than by the purely volumetric $(1 + z)^{3}$. Multiple values of $f_{\textrm{evo}}$ were tested, and a value of $f_{\textrm{evo}}$ approximating the redshift evolution following the models of \cite{Dominguez2011}, \cite{Franceschini2008}, and \cite{Gilmore2012} was selected. The selected value is 1.7. This value does not accurately reproduce the evolution to all redshifts or account for the differences in evolution between the models. The impact on the measurement is discussed in Section~\ref{sec:systematics}. It should be noted that selecting $f_{\textrm{evo}}$ to reproduce the model-based redshift evolution of the EBL introduces a model-dependence into what is otherwise a model-independent measurement.
 
Following the computation of $\tau_{i}(E, z)$ for the 480,000 EBL shapes, the observed photon spectra are corrected for the EBL-absorption predicted by each shape, by multiplying each differential flux point by $e^{\tau(E', z')}$, where $E'$ and $z'$ reflect the energy of the differential flux point and the redshift of the source. A Delaunay interpolation is performed to extract the optical depth at the correct energy and redshift, given that $\tau_{i}(E, z)$ is calculated on a grid in energy and redshift. The numerical uncertainty $\sigma_{i}(\tau_{i}(E, z))$ is propagated into the total uncertainty on the differential flux. The resulting EBL-corrected spectra, $dN_{i}/dE_{i}$, are subjected to a binned $\chi^{2}$ fit. As with the observed spectra, three fit models are considered: a power law, a power law with an exponential cut-off, and a log parabola. In the case that the observed spectrum is well-described by a power law, only a power law is tested for the EBL-corrected spectra, in order to avoid overfitting the data. For EBL-corrected shapes fit with three models, the power law fit is selected if it results in an acceptable $p$-value ($p\geq$~0.05). If a curved spectral shape is required to obtain an acceptable fit, the model with the better $p$-value is selected.

Two assumptions are made about intrinsic blazar spectra, and hence on the possible shapes of the EBL-corrected spectra. The first is that the spectra must be power law or concave: the allowed range on the fit parameters describing curvature is restricted, disallowing convex shapes~\citep{Krennrich2013}. Second, the spectral indices of the EBL-corrected spectra are restricted to be softer (larger) than $\Gamma$=1.0 for the power-law and power-law with exponential cut-off fits. This is a more conservative restriction than the common assumption that the intrinsic spectral index should correspond to that measured by $Fermi$--LAT in the GeV energy range, where EBL attenuation is minimal for the redshifts considered here. The choice of $\Gamma$=1.0 is somewhat arbitrary. It is significantly harder than the average spectral indices measured by $Fermi$--LAT~\citep{3LAC}. However, the second $Fermi$--LAT flare catalog (2FAV) records a small number of flares with $\Gamma\sim$1.5, with large uncertainties~\citep{2FAVA}, motivating the selection of $\Gamma$=1.0.

The EBL shapes are generated with $\lambda_{\textrm{EBL}}$ between 0.1 and 100 $\mu$m, however, the range of $\lambda_{\textrm{EBL}}$ depends on the source redshift and the energy range covered by the photon spectrum. The sensitive wavelength range is defined by the width of the pair-production cross section after integration over the line of sight, and can be approximated as

\begin{equation}
\label{equation:lambdarange}
\lambda_{\rm EBL} \simeq [0.5\,\mu{\rm m} - 5\,\mu{\rm m}] \times \left(\frac{E_\gamma}{1\,{\rm TeV}}\right)\times (1+z)^2,
\end{equation}
taking $E_{\gamma}$ as the energy of the lowest-energy and highest-energy differential flux points in the source's photon spectrum. The highest-energy flux point is taken to be either one flux point beyond the last point with $\geq$2$\sigma$ significance, or the last point with the number of source region counts $>$0 by 2$\sigma$, whichever point is of higher energy.

A projection of the EBL intensity probability distribution at each $\lambda_{\textrm{EBL}}$ grid point is made. Initially a flat distribution, each EBL shape is weighted by exp$(-\chi^{2}$/2), with $\chi^{2}$ taken from the fit to the EBL-corrected spectrum. This choice of weighting disfavors EBL shapes whose EBL-corrected spectra are poorly described by the physically motivated spectral models considered.

\begin{figure}
\centerline{\includegraphics[width=0.5\textwidth]{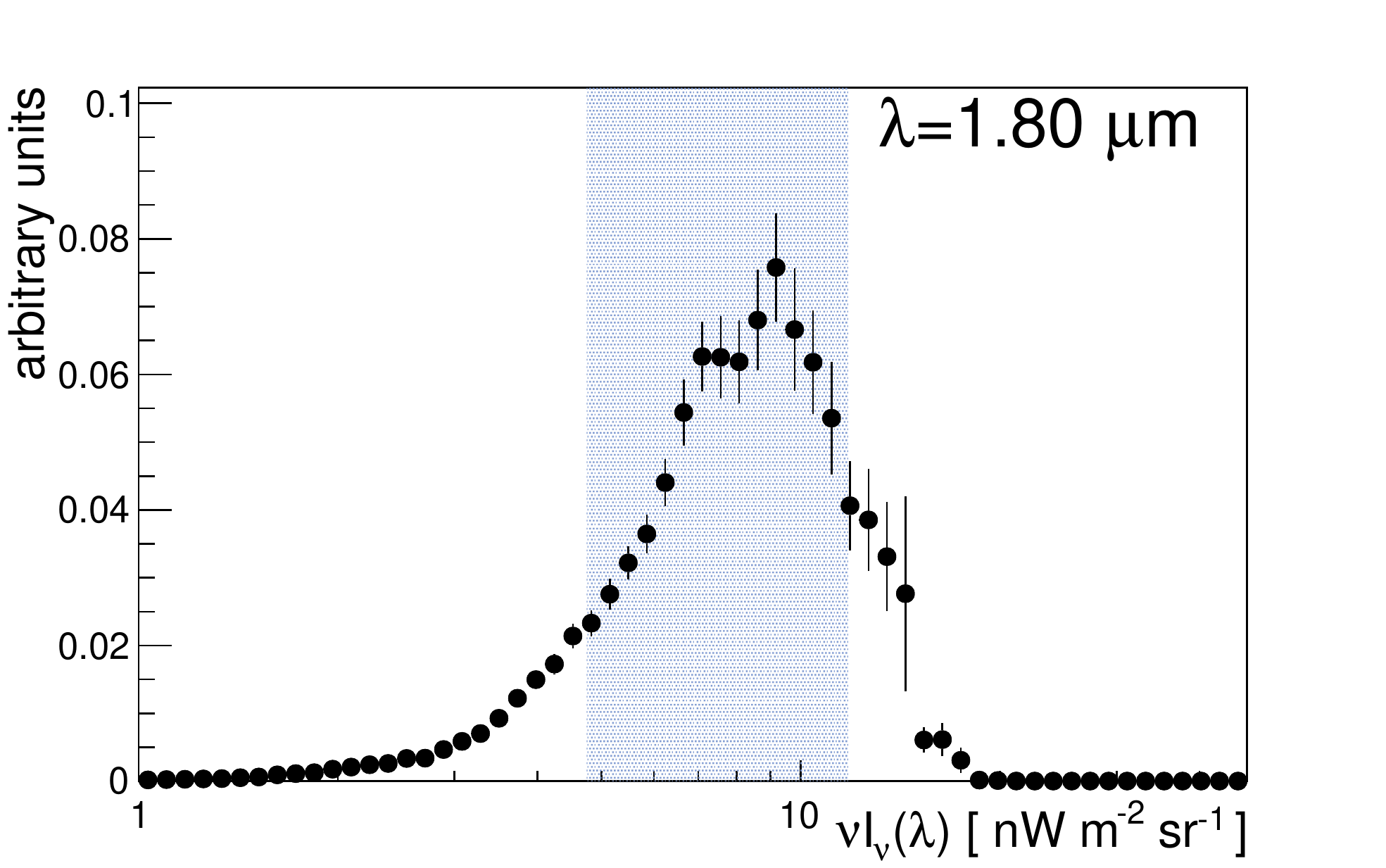}
\includegraphics[width=0.5\textwidth]{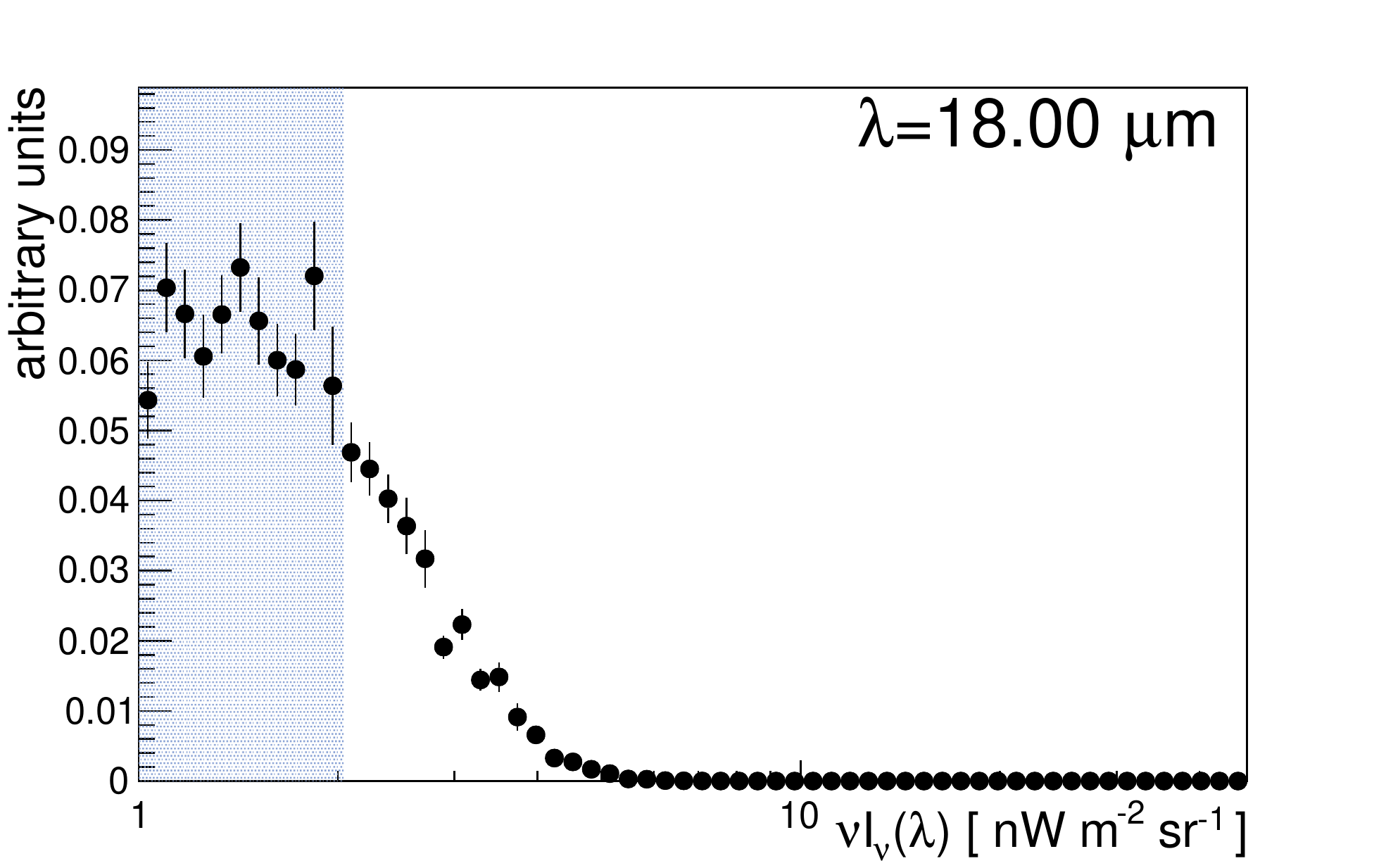}}
\caption{\small{Example EBL intensity probability distributions. The shaded bands indicate the 68\% containment bands. In the left panel, a two-sided containment band for $\nu I_{\nu}(\lambda)$ can be extracted, whereas in the right panel only an upper limit can be set.}}\label{fig:probdenseexample}
\end{figure}

The constraints from multiple blazars are combined by weighting the EBL model by exp$(-(\Sigma\chi^{2})_{n}$/2), where $n$ is an index over the sources. It is worth noting that the spectra are produced with comparable energy binning, ensuring that sources do not receive a greater or lesser weight in the combination based on the choice of binning. At each grid point in $\lambda_{\textrm{EBL}}$, only sources contributing according to Eq.~\ref{equation:lambdarange} are included in the sum. Two example distributions, after $\chi^{2}$ weighting, are shown in Figure~\ref{fig:probdenseexample}. 

A measurement of the EBL intensity is made by integrating the desired (68\% or 95\%) containment. As the probability distributions are in general asymmetric, the quantiles are extracted by integrating inwards from where the distribution tails to zero, rather than outward from a central value. Examples of 68\% containment bands are shown by the shaded regions in Figure~\ref{fig:probdenseexample}. In the right panel, only an upper limit on the EBL intensity can be set, as opposed to the left panel, where both the upper and lower tails of the probability distribution fall to zero.

The accuracy of the method is tested by taking spectra from a bright, nearby source (10 spectra derived from short observations of the Crab Nebula, where event counts are comparable to those of the EBL sources) and calculating the expected spectra at $z$=0.09, 0.2, 0.3, 0.4, 0.5, and 0.6 using the model of \cite{Gilmore2012}. The extracted 68\% containment band contains the input value of the EBL intensity for all redshifts.
 
\section{Systematic Uncertainties}\label{sec:systematics}
Several systematic uncertainties must be considered in the measurement of the EBL SED with the technique described above:
\begin{itemize}
\item{the number of EBL shapes considered,}
\item{the energy scale uncertainty and finite energy resolution of VERITAS,}
\item{uncertainties in the redshift evolution of the EBL,}
\item{the inclusion of sources with uncertain redshifts in the analysis (namely PG 1553+113 and 3C 66A).} 
\end{itemize}

 The choice of 480,000 as the number of generic EBL shapes was made to balance smooth coverage with affordable computing time. Randomly removing 10\% of shapes affected the extracted EBL bounds by less than 1\%, indicating that the precision of measurement is not limited by the number of shapes.
 
The uncertainty on the absolute energy scale of VERITAS is $\sim$20\%~\citep{Madhavan2013}, while the energy resolution is finite and varies with energy. The combination of these two uncertainties propagates into an uncertainty on the photon index $\Gamma$, which becomes relevant when the EBL-corrected spectra are fit with power laws (or power laws with exponential cut-offs) with a limit of $\Gamma$=1.0. A Monte Carlo study was performed per source using the corresponding event counts, effective areas, and exposure times, and assuming a symmetric $\sim$20\% energy scale uncertainty and an energy-dependent energy resolution. The study verified that the energy scale/resolution uncertainties contribute a $\sim$10\% uncertainty on $\Gamma$. By shifting the limiting $\Gamma$ by $\pm 0.10\times\Gamma_{\textrm{obs}}$ and recomputing the containment bands on the EBL intensity, the uncertainty is conservatively accounted for. The maximum impact on the measured EBL intensity is 10\%. 

\begin{figure}
\centerline{\includegraphics[width=0.5\textwidth]{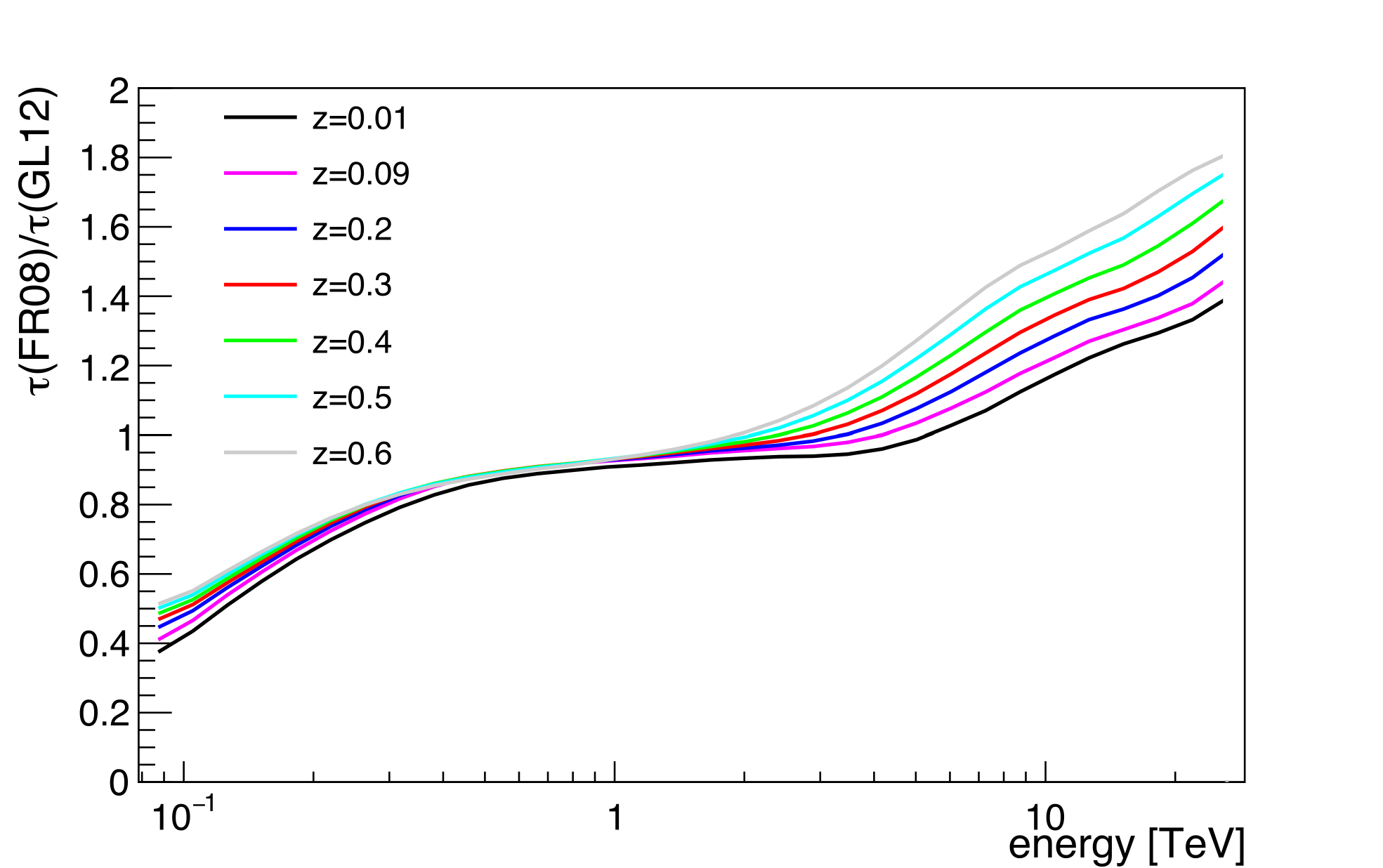}}
\caption{\small{The ratio of $\tau(E)$ predicted by the models of \cite{Franceschini2008} (FR08) and \cite{Gilmore2012} (GL2012).}}\label{fig:fevo_uncertainty}
\end{figure}

As discussed in Sec.~\ref{sec:Analysis}, for the redshift range considered in this analysis, the EBL evolution of \cite{Gilmore2012} is reproduced within uncertainties by using a scaling factor $f_{\textrm{evo}}$=1.7 to treat the redshift evolution of the EBL number density. However, the evolution of the EBL with redshift is not well known, and different models predict different evolution. Figure~\ref{fig:fevo_uncertainty} shows the ratio of $\tau$ predicted by two EBL models, \cite{Franceschini2008} and \cite{Gilmore2012}, as a function of energy for different redshifts. The difference in the predicted values of $\tau$ between the models grows with redshift and energy. The theoretical uncertainty from the spread in the model predictions can be propagated as an additional uncertainty to the EBL SED measurement. The curves shown in Figure~\ref{fig:fevo_uncertainty} are used to obtain an uncertainty as a function of energy by taking the difference in the curve for $z$=0.01 and the curve for the redshift of the source under consideration. The $\tau_{i}(E, z)$ associated with each spectral point of the source's photon spectrum is then varied by the uncertainty, and the most conservative containment band on $\nu I_{\nu}$ is taken. While this does not perfectly account for the spread in the theoretical models---only two models are considered---it attempts to quantify the impact of an uncertainty that has typically been ignored in the literature. The maximum impact on the extracted EBL intensity is 12\%, comparable in size to the impact of the energy scale uncertainty.

Two of the sources included in the measurement, PG 1553+113 and 3C 66A, do not have well-measured redshifts. The redshift ranges for these sources are shown in Table~\ref{sourcelist}. The redshift uncertainty is propagated to the EBL SED measurement as follows: the analysis is performed with the redshifts set to the upper and low bounds of the ranges, and the most conservative containment band on $\nu I_{\nu}$ is taken. 

To summarize, the energy scale uncertainty and the assumed redshift evolution of the EBL are the largest contributors to the total systematic uncertainty. Section~\ref{sec:results} presents the results with and without systematic uncertainties due to the energy scale and redshift evolution. The uncertain redshifts of PG 1553+113 and 3C 66A are however accounted for in all presented results.

\section{Results and Discussion}\label{sec:results}
\begin{figure}
\centerline{\includegraphics[width=0.5\textwidth]{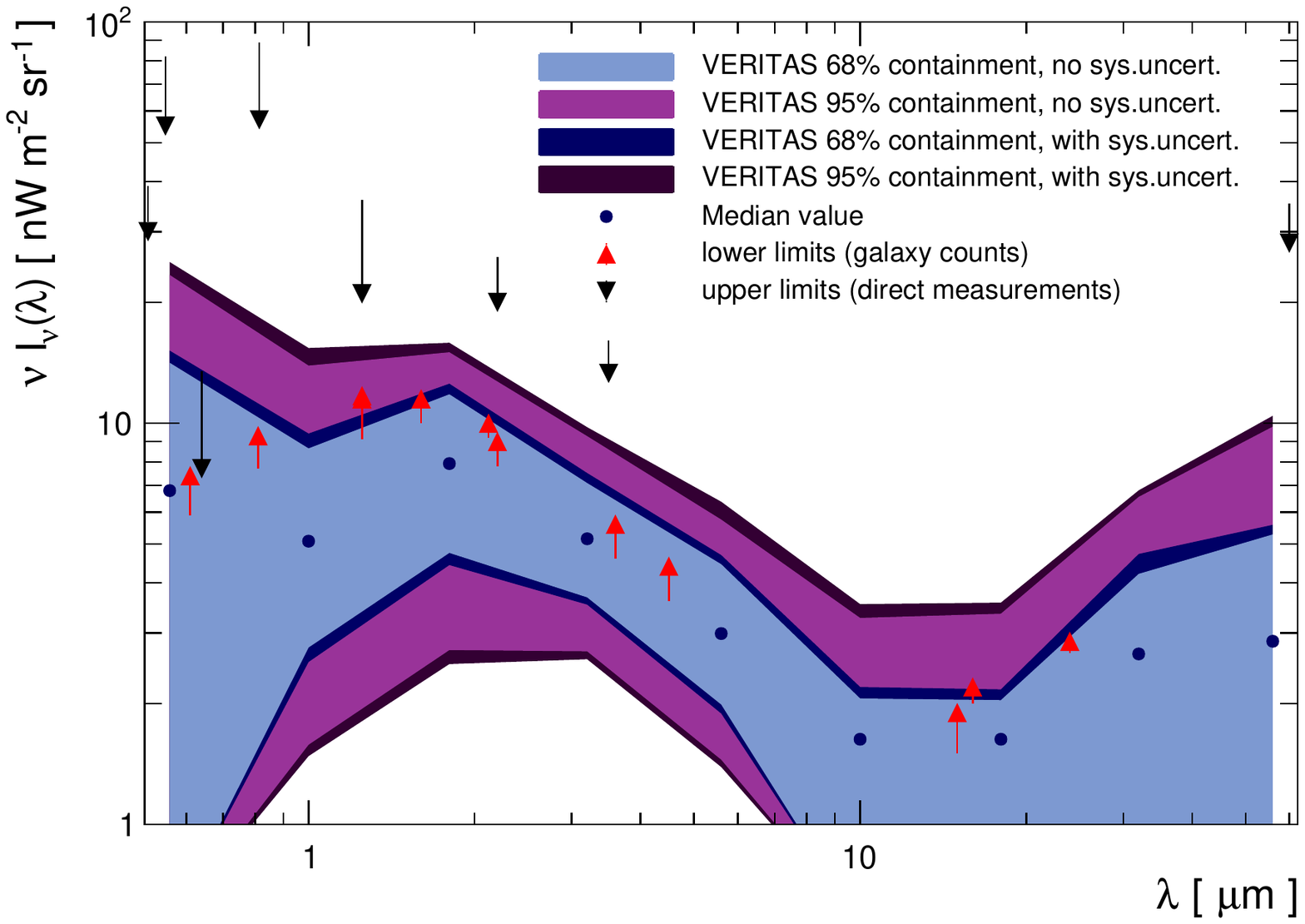}}
\caption{\small{Measured 68\% and 95\% containment bands on the EBL intensity as a function of wavelength. The dark and light shaded regions illustrate the bands with and without systematic uncertainties, respectively. The filled circles mark the median values of the EBL intensity probability distributions. Upward-facing arrows show lower limits from galaxy counts, downward-facing arrows show upper limits from direct measurements.}}\label{fig:SED}
\end{figure}

The measured EBL SED is shown in Figure~\ref{fig:SED}. The 68\% and 95\% containment bands are shown, with the dark and light shaded regions denoting the bands with and without systematic uncertainties, respectively. The filled circles indicate the median values of the EBL intensity probability distributions. Upward-facing arrows indicate the lower limits from galaxy counts, downward-facing arrows show the upper limits from direct measurements (summarized in \cite{Krennrich2013}). The results are further summarized numerically in Table~\ref{tab:EBLresults}, including systematic uncertainties. The measured band is in good agreement with estimates from galaxy counts, indicating no hint of a diffuse component, although the uncertainties are relatively large. The VERITAS results are consistently lower than the limits from direct measurements. 

\begin{table}
\centerline{
\begin{tabular}{ccccc}
$\lambda$   &  $\nu I_{\nu}^{\textrm{min}}$(95\% CI) & $\nu I_{\nu}^{\textrm{min}}$(68\% CI) & $\nu I_{\nu}^{\textrm{max}}$(68\% CI) &  $\nu I_{\nu}^{\textrm{max}}$(95\% CI) \\
$\lbrack~\mu$m$~\rbrack$   & $\lbrack~$nW m$^{-2}$ sr$^{-1}~\rbrack$ & $\lbrack~$nW m$^{-2}$ sr$^{-1}~\rbrack$ & $\lbrack~$nW m$^{-2}$ sr$^{-1}~\rbrack$ & $\lbrack~$nW m$^{-2}$ sr$^{-1}~\rbrack$ \\
\hline
\hline
0.56   &         -    &        -       &  15.2          &   25.2 \\
1.00   &      1.5    &     2.5       &    9.4          &   15.4 \\
1.80   &      2.5    &     4.4       &  12.5          &   15.9 \\
3.20   &      2.6    &     3.5       &    7.5          &   9.8 \\
5.60   &      1.4    &     1.9       &    4.7          &    6.4 \\
10.00 &         -    &        -       &    2.2          &    3.5 \\
18.00 &         -    &        -       &    2.2          &    3.6 \\
32.00 &         -    &        -       &    4.7          &    6.8 \\
56.00 &         -    &        -       &    5.6          &  10.4 \\
 \hline
\end{tabular}
}
\caption{\small{Measured 68\% and 95\% containment bands on the EBL intensity as a function of EBL wavelength, including systematic uncertainties. A dash indicates that no lower limit could be set. The treatment of the redshift uncertainty for two of the sources is discussed in Section~\ref{sec:systematics}.}}
\label{tab:EBLresults}
\end{table}

\begin{figure}
\centerline{\includegraphics[width=0.5\textwidth]{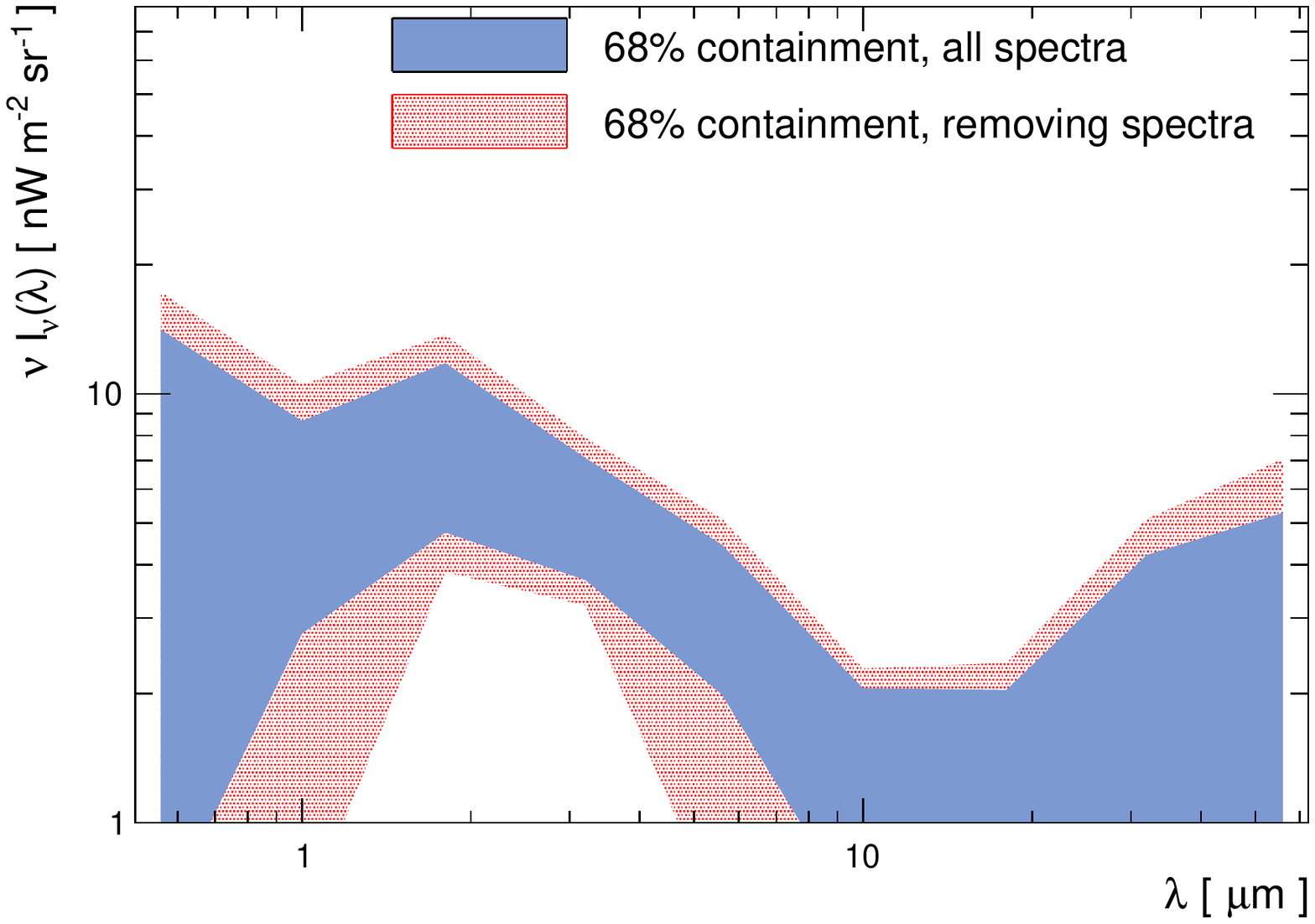}}
\caption{\small{Measured 68\% containment bands on the EBL intensity as a function of EBL wavelength, taking the most conservative bands obtained after removing the sources individually, overplotted with the 68\% containment band including all spectra (without systematic uncertainties).}}\label{fig:SEDjackknife}
\end{figure}

A test of the importance of individual sources was made, shown in Figure~\ref{fig:SEDjackknife}. Each of the sources listed in Table~\ref{sourcelist} was removed from the combined result, and the containment bands recalculated. The most conservative values of $\nu I_{\nu}^{\textrm{min}}$ and $\nu I_{\nu}^{\textrm{max}}$ at each value of $\lambda$ are shown. The resulting 68\% containment band is shown is in Figure~\ref{fig:SEDjackknife}, as is the 68\% containment band considering all spectra (without systematic uncertainties). While the precision of the measurement is reduced when the full source sample is not used, the results are not dominated by any individual source. Indeed, at different wavelengths, the source whose removal most impacts $\nu I_{\nu}^{\textrm{min}}$ and $\nu I_{\nu}^{\textrm{max}}$ varies. The exclusion of the 1ES 1959+650 high state, 1ES 1215+303, 1ES 0229+200, 1ES 1218+304, 1ES 1011+496 and 1ES 0502+675 spectra contribute to the broadening of the containment bands.

\begin{figure}
\centerline{\includegraphics[width=0.5\textwidth]{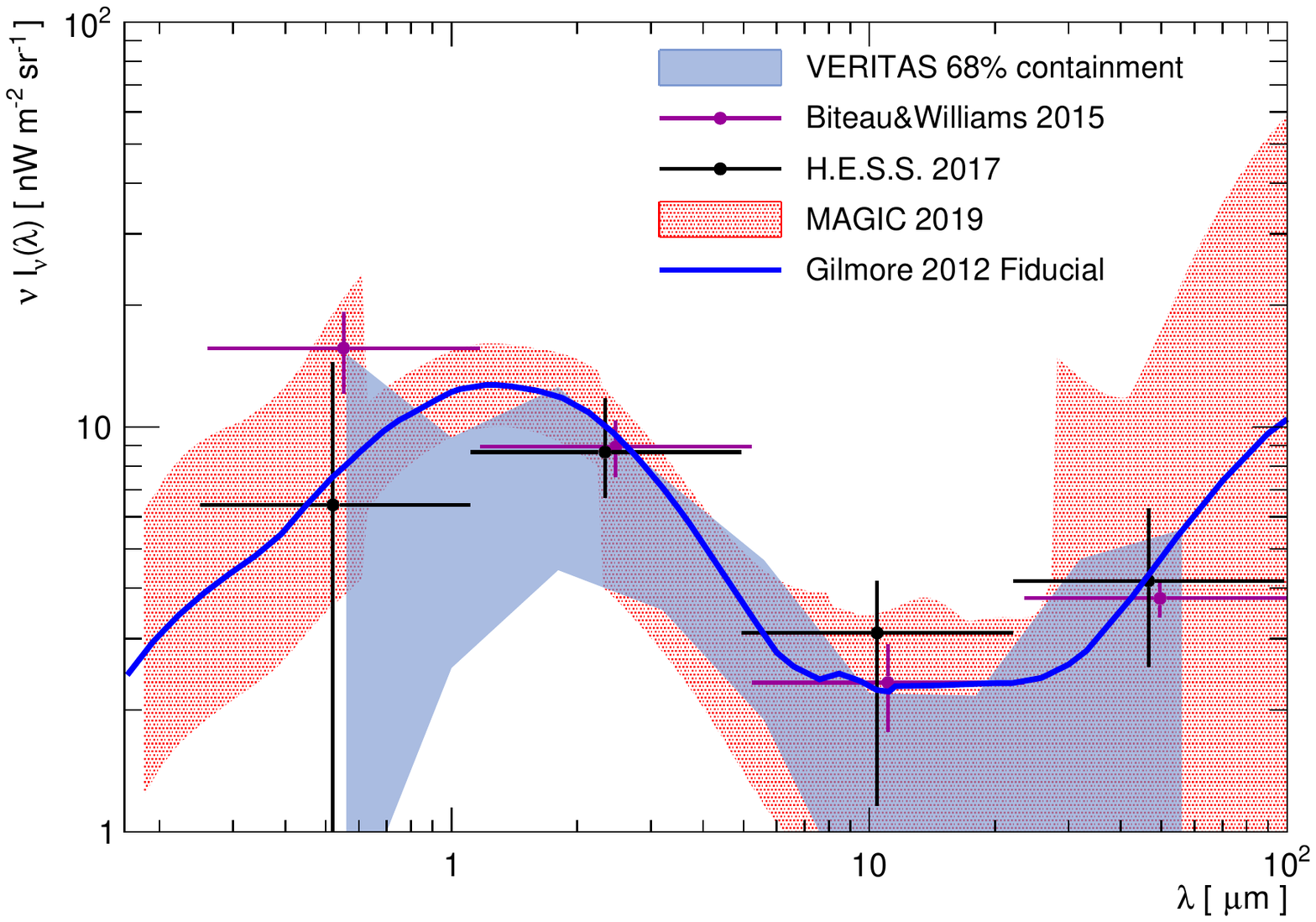}}
\caption{\small{Measured 68\% containment on the EBL intensity as a function of wavelength, compared to the recent gamma-ray measurements of \cite{Biteau2015}, H.E.S.S.~\citep{HESS2017}, and MAGIC~\citep{MAGIC2019}, and the model of \cite{Gilmore2012}.}}\label{fig:SEDcomparison}
\end{figure}

The measurement shown here is in good agreement with other recent measurements. Figure~\ref{fig:SEDcomparison} shows a comparison between the VERITAS 68\% containment band and the results of \cite{Biteau2015} (gamma-ray only, 1$\sigma$ error bars), \cite{HESS2017} (1$\sigma$ error bars, including systematic uncertainties), and \cite{MAGIC2019} (1$\sigma$ error bars, including systematic uncertainties). The model of \cite{Gilmore2012} is included for reference. All gamma-ray measurements are consistent with each other and with \cite{Gilmore2012}, which for this wavelength region is consistent with other state-of-the-art models. Taken together, these measurements point to an EBL SED that is well described by the lower limits from galaxy counts, although the uncertainties remain too large to rule out a diffuse component. It is clearly seen that the approach taken here has more power to constrain high EBL intensities than low EBL intensities, which has been remarked upon in other EBL measurements using imaging atmospheric Cherenkov telescopes~\citep{HESS2017, MAGIC2019}.

\section{Conclusions}
Nine years of VERITAS observations of 14 hard-spectrum blazars, located from redshift $z$=0.044 to $z$=0.604, were used to extract a measurement of the EBL intensity as a function of wavelength. The approach taken made minimal assumptions about the EBL shape and evolution, and made conservative assumptions about the intrinsic spectra of blazars. No individual source is dominant in the resulting joint measurement, and the systematic uncertainties are small in comparison to the statistical uncertainties. This indicates that the VERITAS results can be improved by deeper exposures and better-measured spectra. This is particularly true at large values of $\lambda_{\textrm{EBL}}$, where the state-of-the-art EBL models begin to disagree in their predictions. The agreement between the extracted EBL SED and lower limits from galaxy counts indicates that most, if not all, of the EBL photons can be attributed to resolved galaxies in the wavelength range covered by this measurement. Looking beyond currently operating instruments, a more complete picture of the EBL SED, particularly at the longest wavelengths, will be provided by the forthcoming Cherenkov Telescope Array \citep{CTAscience}. 

%% If you wish to include an acknowledgments section in your paper,
%% separate it off from the body of the text using the \acknowledgments
%% command.
\acknowledgments

This research is supported by grants from the U.S. Department of Energy Office of Science, the U.S. National Science Foundation and the Smithsonian Institution, by NSERC in Canada. This research used resources provided by the Open Science Grid, which is supported by the National Science Foundation and the U.S. Department of Energy's Office of Science, and resources of the National Energy Research Scientific Computing Center (NERSC), a U.S. Department of Energy Office of Science User Facility operated under Contract No. DE-AC02-05CH11231. E.P. acknowledges a Marie Curie Intra-European Fellowship within the 7th European Community Framework Programme and the Young Investigators Program of the Helmholtz Association for support during the period of the project, the DJEI/DES/SFI/HEA Irish Centre for High-End Computing (ICHEC) for the provision of computational facilities and
support, and thanks Jonathan Biteau for useful discussion. We acknowledge the excellent work of the technical support staff at the Fred Lawrence Whipple Observatory and at the collaborating institutions in the construction and operation of the instrument.

%% To help institutions obtain information on the effectiveness of their 
%% telescopes the AAS Journals has created a group of keywords for telescope 
%% facilities.
%
%% Following the acknowledgments section, use the following syntax and the
%% \facility{} or \facilities{} macros to list the keywords of facilities used 
%% in the research for the paper.  Each keyword is check against the master 
%% list during copy editing.  Individual instruments can be provided in 
%% parentheses, after the keyword, but they are not verified.

\vspace{5mm}
\facilities{VERITAS}

\end{document}